\documentclass[aps]{revtex4}

\usepackage{amsmath} 
\usepackage{amsfonts}

\newcommand{\sinlem}{\mathop{\mathrm{sinlem}}}
\newcommand{\arcsinlem}{\mathop{\mathrm{arcsinlem}}}
\newcommand{\coslem}{\mathop{\mathrm{coslem}}}
\newcommand{\sn}{\mathop{\mathrm{sn}}}
\newcommand{\dn}{\mathop{\mathrm{dn}}}

\begin{document}
\title[Maximum attainable energy at a cyclotron]{Maximum attainable energy 
at a fixed frequency cyclotron}

\author{V.V. Konstantinov}
\email{v.konstantinov@g.nsu.ru}
\affiliation{Novosibirsk State University, 630 090, Novosibirsk, Russia}

\author{A. Sen}
\email{abhijit913@gmail.com}
\affiliation{Novosibirsk State University, 630 090, Novosibirsk, Russia}

\author{Z.K.Silagadze}
\email{Z.K.Silagadze@inp.nsk.su}
\affiliation{Budker Institute of Nuclear Physics and Novosibirsk State
University, 630 090, Novosibirsk, Russia.}

\begin{abstract}
We consider a problem of finding the maximum attainable energy at a cyclotron 
as an exercise in the introductory relativity course and comment on some subtle
points of the solution.
\end{abstract}
\maketitle

\section{Introduction}
One of the exercises in an excellent problem book in relativity and gravitation
\cite{1} asks to find out what is the maximum energy one could get out of a
fixed frequency cyclotron with accelerating potential $V$. The solution 
suggested in the book begins with a statement that the energy gain of 
the particle in a cyclotron per acceleration cycle equals to
\begin{equation}
\frac{dE}{dn}=2eV\cos{(\phi-\omega_0 t)},
\label{eq1}
\end{equation}
where $\omega_0$ is the cyclotron frequency and $\phi=\int_0^t \omega dt$ is 
the angular distance traveled by the particle with charge $e$. It is our 
experience that students find (\ref{eq1}) difficult to comprehend. In the 
cyclotron  acceleration takes place at fixed angular locations when the 
particle crosses the narrow gap between two hollow semicircular electrodes, 
called dees. Equation (\ref{eq1}) assumes that the particle's two consecutive 
semicircles in  the dees differ little so that the combined effect of two 
consecutive accelerations is the same as if there were only one accelerating 
gap per cycle with twice bigger accelerating potential $2V$. Then the total 
angular distance $\phi$ traveled by the particle until the next acceleration 
cycle is simply $2\pi n$, $n$ being the number of complete turns accomplished 
by the particle up to a given moment. Therefore $\cos{(\phi-\omega_0 t)}= 
\cos{\omega_0 t}$ and the introduction of $\phi$ into equation (\ref{eq1}) 
seems superfluous. However the remaining part of the solution in \cite{1} 
shows that it is crucial to have $\phi$ in (\ref{eq1}). As a result a student 
stays perplexed. It is our aim in this short note to clarify the issue for 
such a student.

\section{Maximum attainable energy at a cyclotron}
In constant and homogeneous magnetic field $B$ of our idealized cyclotron a 
particle with the charge $e$ and momentum $p$ moves on a circle with the 
radius $R=\frac{pc}{eB}$. Therefore it makes a half-turn in the cyclotron in 
a time $\Delta t=\frac{\pi R}{v}=\frac{\pi mc\gamma}{eB}=\frac{\pi E}{eBc}$, 
$E=mc^2\gamma$ being the particle's energy. During acceleration the energy $E$ 
increases and so does $\Delta t$. As a result for fixed frequency cyclotron it 
becomes impossible to maintain a synchronism in acceleration as the particle 
gets out of phase with accelerating field due to relativistic effects (when 
$\gamma$ begins to appreciably differ from unity). Let's investigate this 
process more closely.

We assume that the particle begins to accelerate in the accelerating gap at 
a time $t_0=0$ in phase with the accelerating potential. Therefore its energy
will become $E_1=E_0+eV$, where $E_0=mc^2$ is the rest energy. The next 
acceleration cycle will take place at a time $t_1=t_0+\frac{\pi E_1}{eBc}$ and 
the particles energy will become $E_2=E_1-eV\cos{\omega_0 t_1}$. The minus
sign reflects the fact that the particle crosses the accelerating gap in the
opposite direction after a half turn (however in the case of perfect 
synchronization the particle still gains energy because the phase of the
accelerating field also changes by $\pi$). The third acceleration cycle
begins at a time $t_2=t_1+\frac{\pi E_3}{eBc}$ and after it the particle's 
energy will become $E_3=E_2+eV\cos{\omega_0 t_2}$. It should be now clear that 
the acceleration process is described by the recurrence relations
\begin{eqnarray} &&
E_m=E_{m-1}+(-1)^{m-1}eV\cos{\omega_0 t_{m-1}},\;\;\; E_0=mc^2,
\nonumber \\ &&
t_m=t_{m-1}+\frac{\pi E_m}{eBc},\;\;\; t_0=0,\; m=1,2,\ldots
\label{eq2}
\end{eqnarray}
To avoid complications related to the alternating-sign factor $(-1)^{m-1}$,
we assume that $eV\ll E_0$. Then two consecutive acceleration cycles will 
differ little and we can embark on the same approximate picture, as in 
(\ref{eq1}), with only one effective accelerating gap per cycle. 
Correspondingly the recurrence relations (\ref{eq2}) will change to
\begin{eqnarray} &&
E_n=E_{n-1}+2eV\cos{\omega_0 t_{n-1}},\;\;\; E_0=mc^2,
\nonumber \\ &&
t_n=t_{n-1}+\frac{2\pi E_n}{eBc},\;\;\; t_0=0,\; n=1,2,\ldots
\label{eq3}
\end{eqnarray}
Repeated application of the second equation of (\ref{eq3}) gives
\begin{equation}
t_n=\frac{2\pi}{eBc}\left (E_n+E_{n-1}+\cdots +E_1\right ),
\label{eq4}
\end{equation}
and the recurrence relation for the energy takes the form
\begin{equation}
E_n=E_{n-1}+2eV\cos{\left [\frac{2\pi \omega_0}{eBc}\left(E_1+E_2+\cdots +
E_{n-1}\right )\right ]}.
\label{eq5}
\end{equation}
Here $n=2,3,\ldots$ and $E_1=E_0+2eV$. Resonance frequency $\omega_0$ of the
cyclotron satisfies $\omega_0\frac{2\pi mc}{eB}=2\pi$, which gives 
$\omega_0=\frac{eBc}{E_0}$. Therefore (\ref{eq5}) can be rewritten as 
\begin{equation}
E_n=E_{n-1}+2eV\cos{\left [\frac{2\pi}{E_0}\left(E_1+E_2+\cdots +
E_{n-1}\right )\right ]}.
\label{eq6}
\end{equation}  
Now we will try to approximately solve this recurrence relation. First of 
all, because $2eV\ll E_0$, we approximate (\ref{eq6}) by the 
integro-differential equation 
\begin{equation}
\frac{dE(n)}{dn}=2eV\cos{\left [\frac{2\pi}{E_0}\int\limits_0^n E(n)\,dn 
\right ]}.
\label{eq7}
\end{equation} 
However (\ref{eq7}) is not quite equivalent to (\ref{eq6}). Namely it doesn't
correctly reproduce the second derivative (second difference) of energy. 
Indeed from (\ref{eq7}) we have 
\begin{equation}
\frac{d^2E(n)}{dn^2}=-\frac{4\pi eVE(n)}{E_0}\sin{\left [\frac{2\pi}{E_0}\int
\limits_0^n E(n)\,dn \right ]}.
\label{eq8}
\end{equation}
While from (\ref{eq6}), by using $\cos{\alpha}-\cos{\beta}=-2\sin{\frac{
\alpha+\beta}{2}}\sin{\frac{\alpha-\beta}{2}}$, we get
\begin{equation}
(E_{n+1}-E_n)-(E_n-E_{n-1})=-4eV\sin{\left [\frac{\pi}{E_0}\left(2E_1+\cdots +
2E_{n-1}+E_n\right )\right ]}\sin{\left (\frac{\pi E_n}{E_0}\right)}.
\label{eq9}
\end{equation}
As we will see soon, the maximal attainable energy will be only slightly 
larger than $E_0$. Therefore we can assume $(E_n-E_0)\ll E_0$. Then
\begin{equation}
\sin{\left (\frac{\pi E_n}{E_0}\right)}=\sin{\left [\pi+\frac{\pi(E_n-E_0)}
{E_0}\right]}=-\sin{\left (\frac{\pi(E_n-E_0)}{E_0}\right )}\approx
-\frac{\pi(E_n-E_0)}{E_0},
\label{eq10}
\end{equation}
and
\begin{eqnarray}
\sin{\left [\frac{\pi}{E_0}\left(2E_1+\cdots +2E_{n-1}+E_n\right )\right ]}
\hspace*{-1mm} = \hspace*{-1mm}\
\sin{\left [\frac{2\pi}{E_0}\left(E_1+\cdots +E_n\right )
\hspace*{-0.5mm} - \hspace*{-0.5mm}\frac{\pi(E_n-E_0)}{E_0} 
\hspace*{-0.5mm} - \hspace*{-0.5mm} \pi\right ]}\hspace*{-1mm}
\approx\hspace*{-0.5mm}
-\sin{\left [\frac{2\pi}{E_0}\left(E_1+\cdots +E_n\right )\right ]}
\hspace*{-0.5mm}.
\label{eq11}
\end{eqnarray} 
Therefore (\ref{eq9}) becomes
\begin{equation}
E_{n+1}-2E_n+E_{n-1}\approx -\frac{4\pi e V(E_n-E_0)}{E_0}
\sin{\left [\frac{2\pi}{E_0}\left(E_1+\cdots +E_n\right )\right ]}.
\label{eq12}
\end{equation}
Clearly we are in trouble, because (\ref{eq8}) and (\ref{eq12}) are not
compatible. However the remedy is simple: we should just add $2\pi n$ to
the argument of cosine in (\ref{eq7}) after which it will take the form
\begin{equation}
\frac{dE(n)}{dn}=2eV\cos{\left [\frac{2\pi}{E_0}\int\limits_0^n (E(n)-E_0)\,dn 
\right ]}.
\label{eq13}
\end{equation}
Of course this addition doesn't change (\ref{eq7}) for integer $n$. But it is
crucial to get the correct second difference (\ref{eq12}), because during
differentiation $n$ no longer remains integer.

Since the particle's full-turn time in the cyclotron equals to $T=\frac{2\pi E}
{eBc}$, its angular velocity is $\omega=\frac{eBc}{E}=\omega_0\frac{E_0}
{E}$. Then $\frac{E}{E_0}dn=\frac{E}{E_0}\,\frac{\omega dt}{2\pi}=
\frac{\omega_0 dt}{2\pi}$ and
\begin{equation}
\frac{2\pi}{E_0}\int\limits_0^n Edn=\int\limits_0^t \omega_0 dt =\omega_0 t,
\;\;\;2\pi\int\limits_0^n dn=\int\limits_0^t \omega dt.
\label{eq14}
\end{equation}
As we see (\ref{eq13}) is fully equivalent to (\ref{eq1}). However now
the origin of the $2\pi n=\int\limits_0^t \omega dt$ term in it is completely 
clear: it is needed to get the second derivative $\frac{d^2 E}{dn^2}$, which 
corresponds to the second difference $E_{n+1}-2E_n+E_{n-1}$, correct. 

Let's denote $q=\frac{dE}{dn}$ and calculate
\begin{eqnarray}
 \frac{dq}{dE}=\frac{dn}{dE}\,\frac{dq}{dn}=-\frac{2eV}{q}\,\frac{2\pi
(E-E_0)}{E_0}\sin{\left [\frac{2\pi}{E_0}\int\limits_0^n (E-E_0)\,dn \right ]}=
-\frac{1}{q}\,\frac{2\pi(E-E_0)}{E_0}\sqrt{(2eV)^2-q^2}.
\label{eq15}
\end{eqnarray}
Obviously, we can separate variables in the obtained differential equation
and obtain
\begin{equation}
d\sqrt{(2eV)^2-q^2}=\frac{\pi}{E_0}d(E-E_0)^2.
\label{eq16}
\end{equation} 
Equation (\ref{eq13}) shows that at the beginning, when $n=0$ and $E=E_0$,
we have $q=2eV$. Hence (\ref{eq13}) can be integrated with the result
\begin{equation}
\sqrt{(2eV)^2-q^2}=\frac{\pi}{E_0}(E-E_0)^2.
\label{eq17}
\end{equation}
The energy $E$ increases so long as $q>0$. Therefore to the maximal energy
there corresponds $q=0$, and then (\ref{eq17}) will give
\begin{equation}
E_{max}=E_0+\sqrt{\frac{2eVE_0}{\pi}}.
\label{eq18}
\end{equation}
As we see, indeed $\frac{E_{max}-E_0}{E_0}=\sqrt{\frac{2eV}{\pi E_0}}\ll 1$,
if $eV\ll E_0$.

\section{Approximate analytical solution}
Equation (\ref{eq18}) solves the problem as formulated in \cite{1}. However
we can go a bit further and provide an approximate analytical solution of
the recurrence relations (\ref{eq6}). From (\ref{eq17}) we have
\begin{equation}
\frac{dE}{dn}=\sqrt{(2eV)^2-\frac{\pi^2}{E_0^2}(E-E_0)^4}.
\label{eq19}
\end{equation}
Therefore
\begin{equation}
n=\int\limits_{E_0}^{E_n}\frac{dE}{\sqrt{(2eV)^2-\frac{\pi^2}{E_0^2}
(E-E_0)^4}}=\sqrt{\frac{E_0}{2\pi e V}}\int\limits_0^{x_n}\frac{dx}
{\sqrt{1-x^4}},
\label{eq20}
\end{equation}
where $x_n=\sqrt{\frac{\pi}{2eVE_0}}(E_n-E_0)$. The final integral is the
simplest case of what are called elliptic integrals. In fact it gives the arc
length of lemniscate, a curve consisting of all points in the plane such that
the product of their distances from two given focal points of the lemniscate 
is constant. In analogy with arc sine function, we can define the lemniscate
sine function and its inverse  by \cite{2}
\begin{equation}
\arcsinlem(x_n)=\int\limits_0^{x_n}\frac{dx}{\sqrt{1-x^4}},\;\;\;
x_n=\sinlem{\left (\int\limits_0^{x_n}\frac{dx}{\sqrt{1-x^4}}\right )}.
\label{eq21}
\end{equation}
Then from (\ref{eq20}) we get
\begin{equation}
E_n=E_0+\sqrt{\frac{2eVE_0}{\pi}}\,\sinlem{\left(\sqrt{\frac{2\pi eV}{E_0}}\,n
\right)}.
\label{eq22}
\end{equation}
From this expression we can find a number of turns required for the particle 
to reach the maximum energy. For lemniscate sine, the number
\begin{equation}
\varpi=2\int\limits_0^1\frac{dx}{\sqrt{1-x^4}}=\frac{\Gamma\left(\frac{1}{4}
\right)^2}{2\sqrt{2\pi}}\approx 2.622
\label{eq23}
\end{equation}
plays the same role as the number $\pi$ for circular trigonometric functions.
In particular lemniscate sine reaches its maximum value $\sinlem{\frac{\varpi}
{2}}=1$ at $\frac{\varpi}{2}$. Therefore it follows from (\ref{eq22}) that
the maximum energy is reached then number of turns $n$ is near to
\begin{equation}
n_m=\frac{\varpi}{2}\sqrt{\frac{E_0}{2\pi eV}}.
\label{eq24}
\end{equation}
For example, if $\frac{eV}{E_0}=10^{-4}$, $n_m\approx 52$.

For small $\frac{eV}{E_0}$, the analytical solution (\ref{eq22}) gives quite
a good approximation for values of $E_n$ calculated from the recurrence 
relations (\ref{eq6}). We have checked up to $n=1000$ that the relative 
accuracy remains better than $10^{-6}$ for $\frac{eV}{E_0}=10^{-4}$, and for 
$\frac{eV}{E_0}=10^{-2}$ the relative accuracy was about $10^{-3}$.    

\section{Concluding remarks}
In our opinion, the considered problem constitutes an excellent exercise in
introductory relativity course. Its solution requires a considerable amount of 
ingenuity and cleverness from the side of student. Besides, in thoughtful 
students this problem can stimulate curiosity in several  interesting 
directions listed below.

\subsection{Vertical focusing and real cyclotrons}
In real cyclotrons one should consider the focusing action of the electric and
magnetic fields, otherwise intensity of the ion beams obtained from the 
cyclotron will be negligible \cite{3,4}.  Inhomogeneities of these fields 
should provide a stabilizing force deflecting the charged particles toward
the median horizontal plane of the acceleration chamber, thus preventing the
particles from spreading in the vertical direction. 

Interestingly, both electric and magnetic vertical focusings were discovered 
expe\-rimentally quite by accident \cite{5,6}. Then a PhD student, 
Stanley Livingston constructed the first cyclotron under Ernest Lawrence's 
guidance. Lawrence thought it was important for the resonance condition to 
have  no electric field inside the dees and thus equipped the accelerating
electrodes with grids at their edges to confine the electric field within
the accelerating gap. Livingston decided to see what would happen without
the grids, as he suspected that the grids intercepted some of the accelerated 
particles thus reducing the beam intensity. In summer, while Lawrence was away 
on a trip, Livingston removed the grids and was surprised to see that the 
resonance was still in place while the  beam intensity increased hundred
times. It was recognized almost immediately that the electric field inside 
the dees produced vertical focusing.

Likewise Livingston empirically discovered that the beam intensity increases
if the magnetic field is a little stronger at the center of the cyclotron 
than at the periphery. Unfortunately such a magnetic field which decreases
radially acts in the same direction as relativity in spoiling the resonance
condition.

After these empirical discoveries it became clear that any discussion of
the cyclotron relativity problem should include beam focusing issues too. 
A careful analysis was performed by Hans Bethe's student Morris Rose \cite{4}.
Bethe's conclusion was disappointing: it would be useless to build cyclotrons
of larger proportions than the existing ones as the relativistic effects will
preclude to reach much higher energies than already obtained. Bethe and Rose 
estimated the maximum  obtainable energy for protons as being of about 12~MeV. 

However Lawrence was not convinced and  responded to Bethe with the remark 
\cite{6}: ``We have learned from repeated experience that there are many ways 
of skinning a cat.'' He believed that money, not relativity, was the main
problem.

A modern cyclotron in Canada (TRIUMF) accelerates protons up to 520~MeV, far
beyond the Bethe-Rose limit. Several key ideas allowed to solve the  
cyclotron relativity problem \cite{7,8,9} and their development constitutes 
a good illustration of the power of human ingenuity, to which Lawrence
faithfully believed.

\subsection{Lemniscate trigonometry}
Generalized sine can be defined \cite{2} as the function inverse to the 
function defined by an integral of the form 
\begin{equation}
x=\int\limits_0^y\frac{dt}{\sqrt{1+m\,t^2+n\,t^4}}.
\label{eq25}
\end{equation}
The case $m=-1$, $n=0$ corresponds to the usual circular sine, $m=1$, $n=0$
gives the hyperbolic sine, and  $m=0$, $n=-1$ corresponds to the lemniscate 
sine encountered in our treatment of the cyclotron relativity problem.
The lemniscate integral has a fascinating history \cite{10,11,12,13,14} 
with the theory of elliptic integrals, elliptic curves and elliptic 
functions \cite{15,16} as a final outcome. The resulting theory is  one of 
the jewels of nineteenth-century mathematics \cite{16}. 

Lemniscate functions in some respects are similar to the trigonometric 
functions but possess certain new characteristics \cite{2}. For example
the analogue of $\sin^2{\phi}+\cos^2{\phi}=1$ in lemniscate case is \cite{2}
$\sinlem^2{t}+\coslem^2{t}+\sinlem^2{t}\,\coslem^2{t}=1$, where $\coslem{t}=
\sinlem{\left(\frac{\varpi}{2}-t\right)}$. 

Usually lemniscate functions are discussed in the more general context of
elliptic functions. Jacobi elliptic functions, found in the mathematical 
description of the motion of a pendulum, are basic elliptic functions. The 
elliptic sine $sn(x,k)$, with modulus $k$, is the inverse of the integral 
(\ref{eq25}) with $m=-(1+k^2)$ and $n=k^2$ for which $1+m\,t^2+n\,t^4=
(1-t^2)(1-k^2\,t^2)$. In terms of $sn(x,k)$ the lemniscate sine is expressed
as follows \cite{2}
\begin{equation}
\sinlem{x}=\frac{1}{\sqrt{2}}\,\frac{\sn\left(\sqrt{2}\,x,\frac{1}{\sqrt{2}}
\right)}{\dn\left(\sqrt{2}\,x,\frac{1}{\sqrt{2}}\right)},
\label{eq26}
\end{equation}
where $\dn(x,k)=\sqrt{1-k^2\sn^2(x,t)}$ is another Jacobi elliptic function.
There exist fast algorithms for computing Jacobi elliptic functions \cite{17}.
So we have used (\ref{eq26}) in our numerical evaluations of  (\ref{eq22}).

Elliptic functions have found numerous applications in physics and it will be
beneficial for students to get acquainted with them. Students may also be
interested in the not very well-known, but fascinating geometry of lemniscate,
associated with various branches of science and mathematics \cite{18}.

\subsection{The standard map and dynamical chaos}
With the identification
\begin{equation}
p_n=\frac{2\pi E_n}{E_0},\;\;\;x_n=\frac{eBc}{E_0}\,t_n+\frac{\pi}{2},\;\;\;
K=\frac{4\pi eV}{E_0},
\label{eq27}
\end{equation}
recurrence relations (\ref{eq3}) take the form
\begin{eqnarray} &&
p_n=p_{n-1}+K\,\sin{x_{n-1}}, \nonumber \\ &&
x_n=x_{n-1}+p_n=x_{n-1}+p_{n-1}+K\,\sin{x_{n-1}}.
\label{eq28}
\end{eqnarray}
In the phase space $(x,p)$, (\ref{eq28}) defines an area-preserving map with
unite Jacobian
$$\frac{\partial(x_n,p_n)}{\partial(x_{n-1},p_{n-1})}=\left |\begin{array}{cc}
1+K\,\cos{x_{n-1}} & 1 \\ K\,\cos{x_{n-1}} & 1\end{array}\right |=1.$$
The map  (\ref{eq28}) is called the Chirikov-Taylor map and it arises in 
a diversity of the physical situations such as kicked rotator (a nonlinear 
oscillator under the influence of periodic kicks-like external force), 
particle dynamics in accelerators, cometary dynamics in celestial mechanics,
charged particles in a magnetic trap, the particle-wave interactions in 
a plasma, Frenkel-Kontorova model, Rydberg atoms ionization  
\cite{19,20,21,22,23}. 

Chirikov called the difference equations (\ref{eq28}) the standard map 
because they model the motion near a nonlinear resonance in any oscillating
system. The standard map and similar area-preserving mappings give rise to 
incredibly rich dynamics and mathematics \cite{23A,23B}. The only parameter of 
the map (\ref{eq28}) is $K$ (the stochasticity parameter) which represents the 
strength of the nonlinearity. Below the critical value $K<K_c\approx 0.97$
\cite{21}, the variation of momentum $p$ is bounded (the curious student may  
learn that this happens due the presence of many  Kolmogorov-Arnold-Moser 
invariant curves and become interested in the mathematically quite demanding
Kolmogorov-Arnold-Moser Theory \cite{24A}). Above the the critical value 
$K>K_c$,  the motion becomes chaotic on large scale, variation of $p$ becomes 
unbounded and the momentum experiences a diffusive growth $p\sim \sqrt{n}$, 
$n$ being the number of iterations. The chaotic component is present even for
small values of $K$, but it is restricted to thin layers of phase space
\cite{24B}. For $K>K_c$ the situation changes to the opposite one, we can 
still find islands of regular motion in the chaotic sea but with the increase
of $K$, their size decreases and for $K\gg 1$ becomes very small \cite{24B}.

One of lessons students can infer from our analysis of the cyclotron relativity
problem is that one should be careful when replacing finite difference 
equations by differential equations or vice verse. In fact such concerns are
of quite a general nature: discretization is equivalent to the introduction of 
a high-frequency external periodic force and in some circumstances a 
considerable difference between the discretized problem and the continuous one
may arise \cite{24}.

We encourage students to investigate (numerically) what happens if the circular
sine in the standard map (\ref{eq28}) is replaced by the lemniscate sine, or if
instead of (\ref{eq28}) one considers the half-standard map, inspired by 
(\ref{eq2}),
\begin{eqnarray} &&
p_n=p_{n-1}+(-1)^{n-1}\frac{K}{2}\,\sin{x_{n-1}}, \nonumber \\ &&
x_n=x_{n-1}+\frac{1}{2}\,p_n.
\label{eq29}
\end{eqnarray}
For small $K\ll 1$, (\ref{eq28}) and (\ref{eq29}) are expected to give nearly 
identical results, but their behavior can differ for a general $K$.


\end{document}